\documentclass[aps,prl,twocolumn,superscriptaddress]{revtex4-2}

\usepackage[T1]{fontenc}
\usepackage{charter}

\usepackage{amsmath,amssymb,amsfonts,graphicx,tabularx,xcolor}

\usepackage[unicode=true,colorlinks=true]{hyperref}

\hypersetup{linkcolor=blue,citecolor=blue,urlcolor=blue}

\begin{document}

\title{Non-Abelian interlayer coherent fractional quantum Hall states}

\author{Xiang-Jian Hou}
\affiliation{School of Physics and Wuhan National High Magnetic Field Center, Huazhong University of Science and Technology, Wuhan 430074, China}

\author{Lei Wang}
\affiliation{National Laboratory of Solid-State Microstructures, Collaborative Innovation Center of Advanced Microstructures, School of Physics, Nanjing University, Nanjing, 210093, China}

\author{Ying-Hai Wu}
\email{yinghaiwu88@hust.edu.cn}
\affiliation{School of Physics and Wuhan National High Magnetic Field Center, Huazhong University of Science and Technology, Wuhan 430074, China}

\begin{abstract}
We study non-Abelian fractional quantum Hall state in double layer systems at total filling factor $1/2$. Recent progresses in two-dimensional van der Waals materials made it possible to explore the regime with very small interlayer distance. Numerical calculations suggests interlayer phase coherence can develop between the layers such that the electrons may redistribute between them without changing the Hall response. It corresponds to spontaneous breaking of the U(1) symmetry associated with the particle number difference in the layers. This state manifests itself as superfluid in counterflow measurement and has characteristic Hall response when current is passed through one layer and voltages in both layers are measured. As the interlayer distance increases, a phase transition to the Halperin 331 state occurs. We also discuss similar physics for bosonic systems with specially designed interactions.
\end{abstract}

\maketitle

{\em Introduction} --- When electrons form Landau levels in two dimensions, kinetic energy is quenched such that the physics is primarily determined by Coulomb interaction in many cases. The striking observations of quantum Hall states initiated the enduring investigations of topological phases~\cite{Halperin2020}. The charged excitations of some fractional quantum Hall (FQH) states are non-Abelian anyons that can be utilized for topological quantum computation~\cite{Nayak2008}. A well-known example of non-Abelian FQH states is the Moore-Read Pfaffian state~\cite{Moore1991}. It was later generalized to the anti-Pfaffian and particle-hole symmetric Pfaffian states based on the analysis of particle-hole symmetry in an isolated Landau level~\cite{Levin2007,LeeSS2007,SonDT2015,Zucker2016,MaKW2024}. Experimental results supporting the Moore-Read type states have been reported in multiple platforms~\cite{Willett1987,PanW1999,Dolev2008,Radu2008,Venka2011,Banerjee2018,Dutta2022a,Dutta2022b,Hossain2018,KiDK2014,KimYW2015,Falson2015,Zibrov2017,LiJIA2017a,ShiQ2020,HuangK2022,HuYW2023,Pack2024,Kumar2024,Singh2024,ChenYW2024,Pandey2024}. 

In double layer quantum Hall systems~\footnote{We use the term double layer instead of bilayer because the latter is reserved to describe bilayer graphene below}, the correlation between two layers have enabled the exploration of excitonic states. At total filling factor $1$, we can perform particle-hole transformation in only one layer such that the electron density in one layer equals the hole density in the other layer. It is possible to generate electron-hole bound states that can be viewed as bosons and form a Bose-Einstein condensate~\cite{Eisenstein2014,Fertig1989,Brey1990,Cote1992,WenXG1992c,EzawaZF1993,YangK1994,MoonK1995,Spielman2000,Spielman2001,Kellogg2002,Kellogg2004,Tutuc2004,Nandi2012}. This system has dramatic experimental ramifications including counterflow superfluidity, interlayer Hall drag, analog of the Josephson effect, and Kosterlitz-Thouless phase transition. Using graphene and transition metal dichalcogenides (TMDs), double layer systems with very small interlayer distance have been fabricated to study exciton condensation~\cite{LiuXM2017,LiJIA2017b,LiuXM2022,ShiQ2022,KimSY2023,LiQX2024}.

This paper studies non-Abelian interlayer coherent (NAIC) FQH states in double layer systems. We choose each layer to be a bilayer graphene (BLG) or a monolayer TMD because multiple even-denominator FQH states observed in them are likely of the Moore-Read type~\cite{Zibrov2017,LiJIA2017a,ShiQ2020,HuangK2022,HuYW2023,Pack2024,Kumar2024}. The interlayer tunneling is negligible such that the numbers of particles in both layers are conserved. When the filling factor of the double layer system is tuned to certain values at which an individual layer supports non-Abelian FQH states, spontaneous interlayer coherence could develop between the two layers under appropriate conditions. The particle number difference between the two layers is spontaneously broken while the total number of particles is still conserved. Numerical calculations are performed to corroborate this picture and identify the suitable range of interlayer distance. Experimental signatures in electric transport and tunneling measurements are analyzed. As the interlayer distance increases, continuous phase transitions eventually occur at some point to destroy the NAIC state. In view of the remarkable experimental progresses, it should be straightforward to verify our theoretical predictions.

\begin{figure}[ht]
\includegraphics[width=0.48\textwidth]{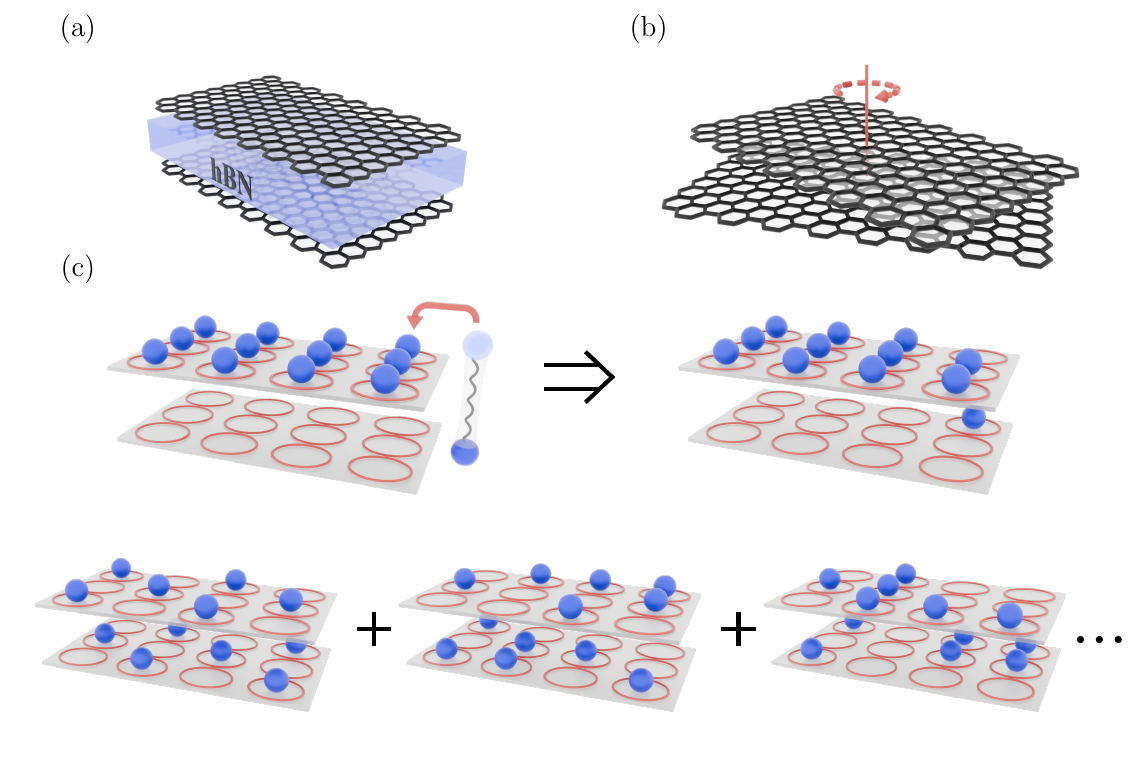}
\caption{Schematics of the double layer system and the exciton injection process. (a) Two layers are separated by a thin layer of hBN. (b) One layer is twisted by a large angle relative to the other. In both panels, one honeycomb lattice is used to represent either bilayer graphene or monolayer TMD. (c) At the beginning, all electrons reside in the top layer and form a Slater determinant. If one exciton is added to the system, one electron in the top layer is annihilated and one electron appears in the bottom layer. The Slater determinant turns into a superposition of many determinants because the excitons may be added to different single-particle states.}
\label{Figure1}
\end{figure}

{\em Model} --- We study fermionic systems in the main text and discuss bosonic systems in Appendix B. The two layers are called top and bottom or pseudospin up and down (denoted as $|\uparrow\rangle$ and $|\downarrow\rangle$). The number of particles in the two layers are $N_{\mathsf{t}}$ and $N_{\mathsf{b}}$. The whole system has filling factor $\nu_{\rm tot}=\nu_{\mathsf{t}}+\nu_{\mathsf{b}}$ with $\nu_{\mathsf{t}}$ ($\nu_{\mathsf{b}}$) being the filling factor of the top (bottom) layer. For our purpose, the interlayer distance $D$ should be smaller than the magnetic length $\ell_{B}=\sqrt{\hbar/(eB)}$. It is important to ensure that the two layers have strong repulsion but negligible tunneling. This is difficult for GaAs quantum wells but can be achieved in van der Waals materials as shown in Fig.~\ref{Figure1} (a-b). In the first approach, two layers of BLG or TMDs are separated by multilayer hexgonal boron nitride (hBN) with thickness down to $2.5$ nm~\cite{LiuXM2017,LiJIA2017b,LiuXM2022}. If we further reduces the thickness, interlayer tunneling might become important. This problem can be averted by the second approach in which the two layers are stacked directly on each other but one layer is twisted by a large angle relative to the other~\cite{ShiQ2022,KimSY2023,LiQX2024}. The valleys in their band structures have very different momenta and this mismatch results in negligible interlayer tunneling at low energy.

The low-energy physics of TMDs is described by the massive Dirac fermion model~\cite{XiaoD2012} whereas that for BLG is more complicated~\cite{JungJ2014}. We consider a simplified model that captures the essential physics in both cases. For non-relativistic Landau levels (LLs), the single-particle eigenstates with orbital index $n$ are denoted as $|\mathsf{L}_{n}\rangle$. In each layer, electrons are confined to one active LL with single-particle eigenstates
\begin{eqnarray}
\begin{bmatrix}
f_{0} |\mathsf{L}_{0}\rangle \\
f_{1} |\mathsf{L}_{1}\rangle
\end{bmatrix}
\label{SingleParticle}
\end{eqnarray}
and mixing with other levels is neglected. It is well-known that the second LL ($n=1$) is favorable for Moore-Read type states and the presence of a small $f_{0}$ would not destabilize them. At moderate magnetic fields, BLG and TMDs correspond to $f_{1} \approx 0.90 \sim 0.99$. The double layer Coulomb potential can be written as 
\begin{eqnarray}
V_{\sigma\tau}(\mathbf{r}_1-\mathbf{r}_2) = \frac{e^2}{4\pi\varepsilon \left[ |\mathbf{r}_1-\mathbf{r}_2|^{2} + (1-\delta_{\sigma\tau})D^{2} \right]^{1/2}}.
\end{eqnarray}
with $\sigma$ and $\tau$ being layer indices. Energy eigenvalues shall be given in units of $e^{2}/(4\pi\varepsilon\ell_{B})$. $N_{\mathsf{t}}$ and $N_{\mathsf{b}}$ are conserved by the microscopic Hamiltonian so the system has U(1)$_{\mathsf{t}}$$\times$U(1)$_{\mathsf{b}}$ symmetry. It can be recast as U(1)$_{+}$$\times$U(1)$_{-}$ that correspond to the conservation of the total number of electrons $N_{e} \equiv N_{\mathsf{t}}+N_{\mathsf{b}}$ and the pseudospin $z$-component $S_{z} \equiv (N_{\mathsf{t}}-N_{\mathsf{b}})/2$. In typical experimental samples, the Coulomb potential would be screened when BLG and TMDs are sandwiched between thick hBN layers. Dynamical polarization effect due to virtual transitions between occupied and empty LLs may also be taken into account~\cite{Zibrov2017}. The fundamental picture to be unveiled below is not likely changed by these modifications as long as single layer Moore-Read type states can be realized.

{\em NAIC state} --- The mechanism of interlayer coherence can be illustrated using the $\nu_{\rm tot}=1$ exciton condensation. As shown in Fig.~\ref{Figure1} (c), the system is initialized to the integer quantum Hall (IQH) state $|\Phi(N_{\mathsf{t}}=N_{e}=N_{\phi},N_{\mathsf{b}}=0)\rangle$ in which all electrons reside in the top layer. We gradually inject excitons to the system where each exciton is formed by one hole in the top layer and one electron in the bottom layer. The filling factor in the top (bottom) layer decreases (increases) but the total filling is intact. The many-body states before and after the injection of one exciton are related via
\begin{eqnarray}
|\Phi(N_{\mathsf{t}}-1,N_{\mathsf{b}}+1)\rangle = \sum_{m} \widehat{C}^{\dag}_{\mathsf{b},m} \widehat{C}_{\mathsf{t},m} |\Phi(N_{\mathsf{t}},N_{\mathsf{b}})\rangle,
\end{eqnarray}
where $\widehat{C}_{\mathsf{t},m}$ ($\widehat{C}^{\dag}_{\mathsf{b},m}$) is the annihilation (creation) operator for the Landau orbital labeled by $m$ in the top (bottom) layer. One electron in the top layer is annihilate so an occupied Landau orbital is vacated to which one electron in the bottom layer is tightly bound. If there is a phase coherence among all injected excitons, the final state would exhibit interlayer coherence as captured by the Halperin 111 wave function~\cite{Halperin1983}. This picture can also be used to construct the NAIC state. We begin with one Moore-Read type state in the top layer
\begin{eqnarray}
\sum_{i} c_{i} |\Phi_{i}(N_{\mathsf{t}}=N_{e}=N_{\phi}/2,N_{\mathsf{b}}=0) \rangle, 
\end{eqnarray}
which is not single Slater determinant but a particular superposition of all possible ones. When excitons are injected to the system, each Slater determinant turns into an interlayer coherent state just as what happened to the IQH state. Meanwhile, the superposition coefficients of the determinants are not changed and inherited by the NAIC state.

\begin{figure}[ht]
\includegraphics[width=0.48\textwidth]{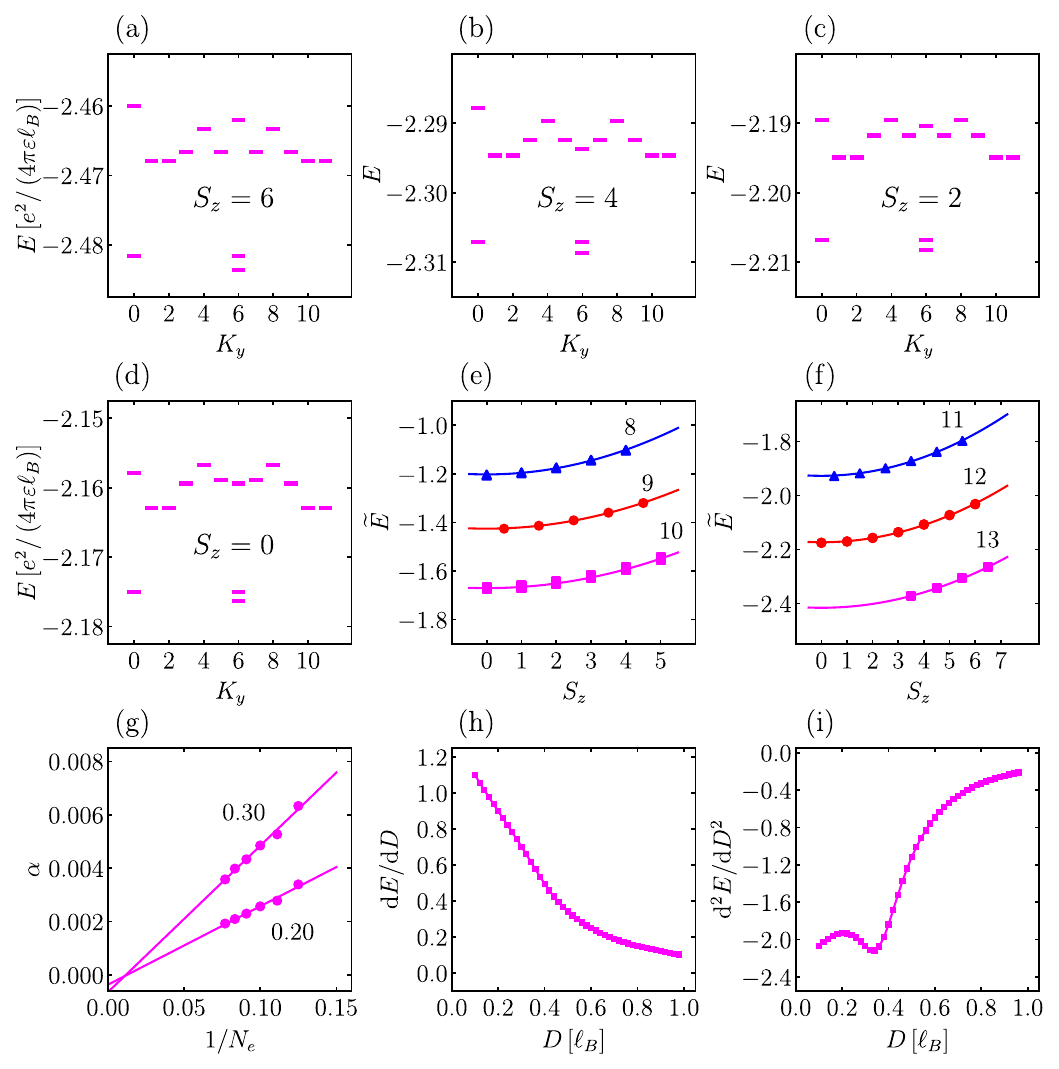}
\caption{(a-d) Energy spectra of the system with $N_{e}=12$ and $D=0.3\ell_{B}$. The $S_{z}$ value is given as inset in each panel. The three quasi-degenerate ground states in all panels have momenta $(K_{x},K_{y}) = (6,0), (0,6), (6,6)$. (e-f) Fitting of the modified eigenvalues $\widetilde{E}(S_{z})$ at $D=0.3\ell_{B}$ using Eq.~\eqref{EnergyFormula}. The number of electrons for each curve is indicated. (g) Finite-size scaling of the coefficient $\alpha$ in Eq.~\eqref{EnergyFormula} at $D=0.2\ell_{B}$ and $0.3\ell_{B}$. (h-i) First- and second-order derivatives of the ground state energy with respect to $D$.}
\label{Figure2}
\end{figure}

To demonstrate that the conjectured state can be realized, we study the microscopic model with $f_{1}=0.95$ and $D=0.3\ell_{B}$ using exact diagonalization on the square torus~\cite{YoshiokaD1983}. At filling factor $\nu_{\rm tot}=p/q$, there is an exact $q$-fold degeneracy due to center of mass translational symmetry. Using relative translational symmetry, two conserved momentum quantum numbers $K_{x}$ and $K_{y}$ can be defined~\cite{Haldane1985b}. For a half-filled system with total magnetic fluxes $N_{\phi}$, both quantities belong to the range $[0,\frac{1}{2}N_{\phi})$. When the Moore-Read type states are realized in single layer systems, the energy spectra for even or odd number of electrons are different. In the former case, three quasi-degenerate ground states appear in the momentum sectors $(K_{x},K_{y})=(\frac{1}{4}N_{\phi},0), (0,\frac{1}{4}N_{\phi}), (\frac{1}{4}N_{\phi},\frac{1}{4}N_{\phi})$~\cite{Peterson2008a,WangH2009}. The total ground state degeneracy (GSD) is six when the center of mass degeneracy is also incorporated. In the latter case, there is only one ground state in the $(0,0)$ sector and the total GSD is two. Low-energy eigenvalues of our double layer system with $N_{e}=12$ and $N_{\phi}=24$ are displayed in Fig.~\ref{Figure2} (a-d). $K_{y}$ is used to as the horizontal axis but $K_{x}$ is suppressed for simplicity. For all possible $S_{z}$ values (only four are shown), three ground states are found in the momentum sectors anticipated for single layer systems. This is strong evidence for the existence of interlayer coherence.

\begin{table}
\begin{tabular}{cc|cc|cc|cc}
\hline
\hline
$N_{e}$ & $K_{y}$ & $S_{z}$ & overlap & $N_{e}$ & $K_{y}$ & $S_{z}$ & overlap\\
\hline
   &   &  5  & 0.9968 &    &   &  5   & 0.9974 \\
   &   &  4  & 0.9793 &    &   &  4   & 0.9802 \\
   &   &  3  & 0.9742 &    &   &  3   & 0.9735 \\
12 & 0 &  2  & 0.9745 & 12 & 6 &  2   & 0.9743 \\
   &   &  1  & 0.9750 &    &   &  1   & 0.9749 \\
   &   &  0  & 0.9716 &    &   &  0   & 0.9697 \\
\hline
   &   & 9/2 & 0.9972 &    &   & 11/2 & 0.9887 \\
   &   & 7/2 & 0.9842 &    &   & 9/2  & 0.9786 \\
11 & 0 & 5/2 & 0.9799 & 13 & 0 & 7/2  & 0.9728 \\
   &   & 3/2 & 0.9763 &    &   &      &      \\
   &   & 1/2 & 0.9802 &    &   &      &      \\
\hline
\hline
\end{tabular}
\caption{Comparison of the exciton injected trial states $|\widetilde{\Psi}(K_{y},S_{z})\rangle$ and exact eigenstates $|\Psi(K_{y},S_{z})\rangle$ at $D=0.3\ell_{B}$.}
\label{Table1}
\end{table}

Next we turn to a more detailed inspection of the ground states and their eigenvalues. The ground state with momentum $K_{y}$ and spin $S_{z}$ is denoted as $|\Psi(K_{y},S_{z})\rangle$. If the state $|\Psi(K_{y},S_{z}+1)\rangle$ is given to us, an exciton injected trial state
\begin{eqnarray}
|\widetilde{\Psi}(K_{y},S_{z})\rangle = \sum_{m} \widehat{C}^{\dag}_{\mathsf{b},m} \widehat{C}_{\mathsf{t},m} |\Psi(K_{y},S_{z}+1)\rangle
\label{TrialState}
\end{eqnarray}
can be generated. If the exciton injection picture is valid, $|\widetilde{\Psi}(K_{y},S_{z})\rangle$ should be a good approximation to the exact eigenstate $|\Psi(K_{y},S_{z})\rangle$. This expectation is confirmed by the overlaps $\langle\widetilde{\Psi}(K_{y},S_{z})|\Psi(K_{y},S_{z})\rangle$ presented in Table~\ref{Table1}. In other words, the exciton operator $\widehat{\psi}^{\dag}_{\rm ex}(x) = \widehat{\psi}^{\dag}_{\mathsf{b}}(x)\widehat{\psi}_{\mathsf{t}}(x)$ exhibits long-range symmetry-breaking order [$\widehat{\psi}_{\sigma}(x)$ is the real space electron annihilation operator]. To compare the eigenvalues from different $S_{z}$ sectors, their bare values should be modified by adding a capacitance energy $dS^{2}_{z}/N_{\phi}$ that arise from the electric charge imbalance~\cite{MacDonald1990}. If $N_{e}$ is odd, the lowest eigenvalue in the $(0,0)$ sector is used. If $N_{e}$ is even, we take the average of the eigenvalues of the quasi-degenerate ground states. In both cases, the modified eigenvalue is denoted as $\widetilde{E}(S_{z})$. For the system with $N_{e}=13$, only four $S_{z}$ sectors have been studied because the Hilbert space dimension gets too large. The data in Fig.~\ref{Figure2} (e-f) can be fitted accurately using
\begin{eqnarray}
\widetilde{E}(S_{z}) = \alpha S^{2}_{z} + \beta.
\label{EnergyFormula}
\end{eqnarray} 
Its coefficient $\alpha$ characterizes the energy scale of pseudospin excitation and extrapolates to zero as the system size increases [see Fig.~\ref{Figure2} (g)]. This means that the system supports gapless excitations associated with spontaneous breaking of the U(1)$_{-}$ symmetry. The quadratic dependence in Eq.~\eqref{EnergyFormula} is reminiscent of the excitation spectrum in XY easy-plane ferromagnet, which was also found in the $\nu_{\rm tot}=1$ IQH exciton condensate~\cite{Eisenstein2014}. While it is natural to assign definite layer index (pseudospin $z$ component) to each electron, quantum superpositions $|\uparrow\rangle+\exp(i\phi)|\downarrow\rangle$ with an arbitrary phase $\phi$ are also legitimate choices. If all electrons are placed in one such pseudospin state, the system becomes a ferromagnet in the XY plane. The value of $\phi$ is uniform in the ground state and its variation in space generates gapless excitations.

{\em Experimental signatures} --- Electric transport measurement is useful for probing the NAIC state. For both Hall bar and Corbino geometry, three measurement schemes could be utilized to study different aspects~\cite{Kellogg2002,Kellogg2004,Tutuc2004,LiuXM2017,LiJIA2017b,LiuXM2022}. In parallel flow measurement, the system behaves as an ordinary FQH state. In counter flow measurement, the system is highly conductive (and should behave like a superfluid in the ideal limit). In drag measurement, a current $I_{\sigma}$ is passed through the layer $\sigma$ and voltages $V_{\mathsf{t}/\mathsf{b}}$ in both layers are measured. The results are summarized in the Hall resistance matrix
\begin{eqnarray}
\mathbf{R}_{xy} = \begin{bmatrix}
R^{\mathsf{tt}}_{xy} & R^{\mathsf{tt}}_{xy} \\
\\
R^{\mathsf{bt}}_{xy} & R^{\mathsf{bb}}_{xy}
\end{bmatrix}
\label{HallMatrix}
\end{eqnarray}
with $R^{\sigma\tau}_{xy}=V_{\tau}/I_{\sigma}$. Its diagonal (off-diagonal) elements are called drive (drag) Hall resistance. The inverse of $\mathbf{R}_{xy}$ is the Hall conductance matrix that is more convenient to study from the theoretical perspective. The elements of this matrix are many-body Chern numbers defined in the space of twisted boundary angles~\cite{NiuQ1985}. In particular, off-diagonal elements may be obtained when the top (bottom) layer has nonzero twist angle along the $x$ ($y$) direction (or the other way around)~\cite{ShengDN2003}. This method is not viable in our system because it is gapless in the counterflow channel. Instead, we employ the argument proposed in Ref.~\cite{YangK1998} to deduce that all elements in Eq.~\eqref{HallMatrix} are $2h/e^{2}$. This means that the Hall resistance matrix is singular, so it cannot be obtained using twisted boundary conditions. 

Another important signature is about tunneling between the two layers~\cite{Spielman2001}. We consider a process in which one electron changes its layer index but remains at the same spatial location, as described by the operator $H_{\rm tun} \sim \widehat{\psi}^{\dag}_{\mathsf{t}}(x) \widehat{\psi}_{\mathsf{b}}(x) + \widehat{\psi}^{\dag}_{\mathsf{b}}(x) \widehat{\psi}_{\mathsf{t}}(x)$~\cite{ZhangYH2017} . The tunneling current-energy relation is encoded in the spectral function
\begin{eqnarray}
I(\omega) = \sum_{n} \left| \langle \mathsf{M}_{n} | \widehat{\psi}^{\dag}_{\mathsf{b}}(x) \widehat{\psi}_{\mathsf{t}}(x) | \mathsf{M}_{0} \rangle \right|^{2} \delta(\omega-E_{n}+E_{0}),
\end{eqnarray}
where $|\mathsf{M}_{n}\rangle$ is the $n$-th many-body eigenstate with energy $E_{n}$ ($n=0$ is the ground state). An expansion of the real space operators in the active LL yields $\widehat{\psi}^{\dag}_{\mathsf{b}}(x) \widehat{\psi}_{\mathsf{t}}(x) = \sum_{m} g_{m} \widehat{C}^{\dag}_{\mathsf{b},m} \widehat{C}_{\mathsf{t},m}$ with certain coefficients $g_{m}$. It is easy to see that $\mathrm{d}I(\omega)/\mathrm{d}\omega$ has a zero bias peak due to the following two properties: the eigenstates in two neighboring $S_{z}$ sectors are related via Eq.~\eqref{TrialState} and the system is gapless in the thermodynamic limit ($E_{1}-E_{0} = \alpha$ that approaches zero). An exceptional problem about the Moore-Read type states is how to distinguish between Pfaffian, anti-Pfaffian, and particle-hole symmetric Pfaffian~\cite{Moore1991,Levin2007,LeeSS2007,SonDT2015,Zucker2016}. It is intimately connected with the presence and breaking of particle-hole symmetry in one LL. The precise nature of a Moore-Read type state can be deduced from experimental results about thermal transport~\cite{Banerjee2018,Dutta2022a,Dutta2022b}, daughter states~\cite{Levin2009,Yutushui2024a,Zhelton2024}, chiral gravitons~\cite{LiouSF2019,Haldane2021,NguyenDX2021a,LiangJH2024}, and shot noise~\cite{ParkJH2020,Yutushui2022,Manna2024}. For our double layer model, particle-hole symmetry maps a state at $\nu_{\rm tot}$ to its counterpart at $2-\nu_{\rm tot}$ (rather than at $1-\nu_{\rm tot}$). Interlayer coherent versions of the daughter states can be defined, and some of them may be stabilized for appropriate interlayer distance and charge imbalance. We also discuss briefly chiral gravitons of the NAIC state in Appendix A.

{\em Phase transitions} --- When the interlayer distance increases, other states are expected to arise so there may be nontrivial phase transitions. For the exciton condensate at $\nu_{\rm tot}=1$, similar questions have been extensively studied and the results are highly debated~\cite{ParkK2004,Moller2009,Alicea2009,Milovan2015,ZhuZ2017b,LianB2018,Wagner2021,Ruegg2023,DengHY2024,Lotric2024,Ruegg2024}. The physics generally depend on charge distribution in the layers and we first study the balanced cases. In the limit of infinite $D$, two layers are decoupled and each one realizes a $\nu=1/4$ composite fermion liquid~\cite{Halperin1993,Rezayi2000}. For the intermediate $D$ regime, a natural candidate is the Halperin 331 state~\cite{Halperin1983} with the GSD being eight. Besides the center of mass degeneracy, there is one state in each of the momentum sectors $(K_{x},K_{y})=(0,0), (\frac{1}{4}N_{\phi},0), (0,\frac{1}{4}N_{\phi}), (\frac{1}{4}N_{\phi},\frac{1}{4}N_{\phi})$. This is indeed observed at several $D \sim 0.6\ell_{B}$. By tuning the interlayer distance, there must be a phase transition and its properties are quite intriguing. As a first step, we may study the evolution of the ground state energy. However, it is not entirely clear how to define this quantity. There are multiple quasi-degenerate ground states on the torus and the degeneracy changes in the transition. We take the average of the six quasi-degenerate ground states of the NAIC state. The first- and second-order derivatives of this average are presented in Fig.~\ref{Figure2} (h-i). One may tentatively claim that a continuous quantum phase transition occurs at $D \approx 0.34 \ell_{B}$. There should be another transition to the decoupled CFL but it is quite difficult to identify based on existing numerical results. Besides these zero temperature transitions, finite temperature transitions are also expected. The excitonic correlation in our system renders superfluid like behavior and may be destroyed in a Kosterlitz-Thouless type transition. If we turn to the charge imbalanced cases, the physics could be even more complicated. When the system is extremely imbalanced ($S_{z} \sim N_{e}/2$), the number of excitons is small and coherence among them may persist in a larger range of $D$. In other words, there may be no two-component FQH states to compete with the NAIC state for certain charge distributions.

{\em Discussions and conclusions} --- To search for the NAIC state, one should be aware of the difference between the actual filling factor $\nu_{\rm act}$ in experimental samples and the variable $\nu_{\rm tot}$ defined above. The former is measured with respect to the charge neutral point whereas the latter is concerned with the active LL alone (partially occupied by electrons). Previous works have observed FQH states in BLG at $\nu_{\rm act}=-5/2,-3/2,-1/2,1/2,3/2,5/2,7/2$~\cite{Zibrov2017,LiJIA2017a,HuangK2022,HuYW2023,Kumar2024}, whose associated NAIC states in double BLG have $\nu_{\rm act}=-11/2,-7/2,-3/2,1/2,5/2,9/2,13/2$. There is one additional subtle point: a suitable perpendicular electric field is required to generate some states in BLG. If such a field is also applied in the double layer case, the single-particle energy of the two layers are shifted such that the gapless excitations may be gapped out. This suggests that NAIC states are more likely found at $\nu_{\rm act}=-7/2,-3/2,9/2,13/2$.

An effective field theory for the NAIC state is desirable. The $\nu_{\rm tot}=1$ exciton condensate is described by an Abelian Chern-Simons-Maxwell theory with two gauge fields: one for the parallel flow quantized Hall response and one for the counterflow superfluid~\cite{WenXG1992c}. Naively, the former one should be replaced by a non-Abelian gauge field in our system~\cite{Fradkin1998,WenXG1999}. If this construction is successful, one may proceed to construct a field theory for the phase transition to the Halperin 331 state, which is described by an Abelian Chern-Simons theory~\cite{WenXG1992a}. This process involves an conversion from non-Abelian to Abelian gauge fields in the charge sector and gapping out the Chern-Simons-Maxwell terms in the pseudospin sector. We discuss this issue for the bosonic NAIC state in Appendix B. 

It would also be interesting to explore similar physics for the non-Abelian Read-Rezayi $\mathbb{Z}_{3}$ state at $\nu=3/5$~\cite{Read1999}, whose particle-hole conjugate is a good candidate for the $\nu=2+2/5$ state in GaAs~\cite{Rezayi2009,ZhuW2015c,MongR2017}. Multiple states at filling factors $p+2/5$ and $p+3/5$ (with integer $p$) have been observed in BLG and TMD, but their non-Abelian nature is difficult to ascertain. We expect that NAIC Read-Rezayi states can only be found in specially designed systems. In particular, the interlayer distance should also be small, so BLG and TMD are still more preferable than GaAs in this regard. The inclusion of additional internal degrees of freedom such as spin or valley may lead to more diverse phenomena~\cite{Pellegrini1998,DasSarma1998,MacDonald1999,Brey1999,Saha2022}.

{\em Acknowledgements} --- Some calculations were performed using the DiagHam package for which we are grateful to the authors~\cite{DiagHam}. X. J. H. and Y. H. W. were supported by National Natural Science Foundation of China (Grant No. 12174130). L. W. was supported by the National Key Projects for Research and Development of China (Grant Nos. 2021YFA1400400, 2022YFA1204700) and Natural Science Foundation of Jiangsu Province (Grant No. BK20220066).

\bibliography{ReferConde}

\clearpage
\onecolumngrid

\setcounter{figure}{0}
\setcounter{table}{0}
\setcounter{equation}{0}
\renewcommand{\thefigure}{A\arabic{figure}}
\renewcommand{\thetable}{A\arabic{table}}
\renewcommand{\theequation}{A\arabic{equation}}

\appendix

\section{Appendix A: Chiral Gravitons}

This section studies chiral gravitons of the NAIC state. To establish the framework, we begin with non-relativistic Landau levels~\cite{LiouSF2019,Haldane2021,NguyenDX2021a}. The single-particle Hamiltonian is
\begin{eqnarray}
\mathcal{H} = \frac{1}{2\mu} \left[ \Lambda \widehat{p}^{2}_{x} + \frac{1}{\Lambda} \left( \widehat{p}_{y} - eBx \right)^{2} \right],
\end{eqnarray}
where $B$ is the magnetic field and $\Lambda$ is the mass anisotropy. The plane is wrapped to a rectangular torus spanned by the vectors $\mathbf{L}_{x} = L_{x} \mathbf{e}_{x}$ and $\mathbf{L}_{y} = L_{y} \mathbf{e}_{y}$. The number of fluxes through the torus is $N_{\phi}$ and it satisfies the constraint $2\pi\ell^{2}_{B}N_{\phi}=L_{x}L_{y}$. A momentum variable is written generally as $\mathbf{q}=q_{x}\mathbf{G}_{x}+q_{y}\mathbf{G}_{y}$ using the reciprocal vectors $\mathbf{G}_{x} = 2\pi\mathbf{e}_{x}/L_{x}$ and $\mathbf{G}_{y} = 2\pi\mathbf{e}_{y}/L_{y}$. For electrons confined to the $\alpha$-th LL, the many-body Hamiltonian is
\begin{eqnarray}
H(\Lambda) && = \frac{1}{2L_{x}L_{y}} \sum_{\sigma,\tau} \sum_{\{m_{i}\}} \sum_{q_{x},q_{y}} \left[ \mathcal{L}_{\alpha}\left( Q^{2}_{\Lambda}/2 \right) \right]^{2} \; V(\mathbf{q}) \exp \left[ -\frac{1}{2} Q^{2}_{\Lambda} \ell^{2}_{B} - i\frac{2{\pi}q_{x}}{N_{\phi}} \left( m_{1}-m_{4} \right) \right] \nonumber \\
&& \times \widetilde{\delta}_{m_{1},m_{3}-q_{y}} \widetilde{\delta}_{m_{2},m_{4}+q_{y}} \; \widehat{C}^{\dag}_{\sigma,m_{1}} \widehat{C}^{\dag}_{\tau,m_{2}} \widehat{C}_{\tau,m_{4}} \widehat{C}_{\sigma,m_{3}}.
\label{AnisoHamilton}
\end{eqnarray}
Here $V(\mathbf{q})$ is Fourier transform of the interaction potential,
\begin{eqnarray}
Q^{2}_{\Lambda} = \Lambda \left( \frac{2\pi\ell_{B}}{L_{x}} q_{x} \right)^{2} + \frac{1}{\Lambda} \left( \frac{2\pi\ell_{B}}{L_{y}} q_{y} \right)^{2},
\end{eqnarray}
$\widetilde{\delta}_{s,t}$ is a generalized Kronecker symbol that equals one if $s=t$ up to multiples of $N_{\phi}$, and $\mathcal{L}_{\alpha}$ is the $\alpha$-th Laguerre polynomial.

We introduce time-dependent anisotropy $\Lambda=1+\xi(t)$ to excite chiral gravitons. The Hamiltonian can be expanded as
\begin{eqnarray}
H(\Lambda) &=& H(1) + \xi(t) \mathcal{O}
\end{eqnarray}
with
\begin{eqnarray}
\mathcal{O} && = \frac{1}{2L_{x}L_{y}} \sum_{q_{x},q_{y}} \sum_{\{m_{i}\}} -\frac{1}{2} \left[ \left( \frac{2\pi\ell_{B}}{L_{x}} q_{x} \right)^{2} - \left( \frac{2\pi\ell_{B}}{L_{y}} q_{y} \right)^{2} \right] \left\{ \left[ \mathcal{L}_{\alpha}\left( Q^{2}_{1}/2 \right) \right]^{2} - 2 \mathcal{L}_{\alpha}\left( Q^{2}_{1}/2 \right) \mathcal{L}'_{\alpha}\left( Q^{2}_{1}/2 \right) \right\} \nonumber \\
&& \times V(\mathbf{q}) \exp \left[ -\frac{1}{2} Q^{2}_{1} \ell^{2}_{B} - i\frac{2{\pi}q_{x}}{N_{\phi}} \left( m_{1}-m_{4} \right) \right] \widetilde{\delta}_{m_{1},m_{3}-q_{y}} \widetilde{\delta}_{m_{2},m_{4}+q_{y}} \; \widehat{C}^{\dag}_{\sigma,m_{1}} \widehat{C}^{\dag}_{\tau,m_{2}} \widehat{C}_{\tau,m_{4}} \widehat{C}_{\sigma,m_{3}}.
\label{SimpleGravi}
\end{eqnarray}
This operator does not reveal the graviton chirality, so we perform the replacement
\begin{eqnarray}
\left( \frac{2\pi\ell_{B}}{L_{x}} q_{x} \right)^{2} - \left( \frac{2\pi\ell_{B}}{L_{y}} q_{y} \right)^{2} \quad \rightarrow \quad \left( \frac{2{\pi}}{L_{x}}q_{x} \mp i \frac{2{\pi}}{L_{y}}q_{y} \right)^{2}
\end{eqnarray}
to define chiral graviton operators $\mathcal{O}_{\pm}$. The eigenvalues and eigenstates of $H(1)$ are denoted as $E_{n}$ and $|\mathsf{M}_{n}\rangle$ ($n=0$ is the ground state). We are interested in the spectral functions
\begin{eqnarray}
I_{\pm} = \frac{1}{W_{\pm}} \sum_{n} \left| \langle \mathsf{M}_{n} | \mathcal{O}_{\pm} | \mathsf{M}_{0} \rangle \right|^{2} \delta(\omega - E_{n} + E_{0})
\end{eqnarray}
with $W_{\pm} = \langle \mathsf{M}_{0} | \left[ \mathcal{O}_{\pm} \right]^{\dag} \mathcal{O}_{\pm} | \mathsf{M}_{0} \rangle$ being the total weight.

When spinor eigenstates in Eq.~\eqref{SingleParticle} are used to describe BLG and TMD, the many-body Hamiltonian and chiral graviton operators should be modified. The $\alpha$-th Laguerre polynomial in Eq.~\eqref{AnisoHamilton} is changed to 
\begin{eqnarray}
|f_{0}|^{2} + |f_{1}|^{2} \mathcal{L}_{1}\left( Q^{2}_{\Lambda}/2 \right) = |f_{0}|^{2} + |f_{1}|^{2} \left( 1- Q^{2}_{\Lambda}/2 \right)
\end{eqnarray}
and the factor inside the braces in Eq.~\eqref{SimpleGravi} is changed to
\begin{eqnarray}
&& \left[ |f_{0}|^{2} + |f_{1}|^{2} \mathcal{L}_{1}\left( Q^{2}_{1}/2 \right) \right]^{2} - 2 |f_{1}|^{2} \left[ |f_{0}|^{2} + |f_{1}|^{2} \mathcal{L}_{1}\left( Q^{2}_{1}/2 \right) \right] \mathcal{L}'_{1}\left( Q^{2}_{1}/2 \right) \nonumber \\
= && \left[ |f_{0}|^{2} + |f_{1}|^{2} \left( 1- Q^{2}_{\Lambda}/2 \right) \right] \left[ |f_{0}|^{2} + 2|f_{1}|^{2} + |f_{1}|^{2} \left( 1- Q^{2}_{\Lambda}/2 \right) \right].
\end{eqnarray}
It has been found that the dominant chirality is negative (positive) in the Pfaffian (anti-Pfaffian) state whereas both chiralities are equal in the particle-hole symmetric Pfaffian state~\cite{Haldane2021}. For an isolated LL with two-body interaction, particle-hole symmetry ensures that $W_{+}=W_{-}$. This occurs in our model when all electrons reside in the same layer ($S_{z}=N_{e}/2$). If there are electrons in both layers, particle-hole symmetry instead relates filling factors $\nu_{\rm tot}$ and $2-\nu_{\rm tot}$. We have computed the spectral functions for the system with $N_{e}=8$. As the charge imbalance decreases, the ratio $W_{-}/W_{+}$ slightly increases to $1.19$ at $S_{z}=0$. This may be taken as evidence that the NAIC state is of the Pfaffian type, but it is still premature to make such a claim. For experimental systems, LL mixing is expected to have a much large impact on $W_{-}/W_{+}$ than the effect observed here. A more detailed investigation is left for future works.

\section{Appendix C: Bosonic Systems}

This section studies NAIC state in bosonic systems. The results not only demonstrate the generality of exciton physics but also provides insights into the field theory description. It is well established that the bosonic Pfaffian state at filling factor $1$ can be realized~\cite{Cooper2001,Regnault2003,ChangCC2005}. Instead of the Coulomb potential, we parameterize the interaction potential between bosons using the Haldane pseudopotentials~\cite{Haldane1983c}. In a double layer system, the pseudopotential terms are written as $V^{i}_{\sigma\tau}$, where $i$ is the relative angular momentum of two bosons and $\sigma,\tau$ are layer indices. We study a double layer system described by the Hamiltonian
\begin{eqnarray}
H_{\rm boson} = V^{0}_{\mathsf{tt}} + 1.5 V^{1}_{\mathsf{tt}} + 0.2 V^{2}_{\mathsf{tt}} + V^{0}_{\mathsf{bb}} + 1.5 V^{1}_{\mathsf{bb}} + 0.2 V^{2}_{\mathsf{bb}} + g \left( V^{0}_{\mathsf{tb}} + 1.5 V^{1}_{\mathsf{tb}} + 0.2 V^{2}_{\mathsf{tb}} \right)
\end{eqnarray}
at total filling factor $\nu_{\rm tot}=1$. For bosons in the same layer, the zeroth and second pseudopotentials are used to produce robust numerical signatures of the Pfaffian state (the first pseudopotential is irrelevant). All three terms are useful when two bosons reside in different layers. In a realistic setting, pseudpotentials are expected to decay as the relative angular momentum increases. However, we need to choose the non-monotonic coefficients $1.0,1.5,0.2$ to stabilize the bosonic NAIC state.

\begin{figure}[ht]
\includegraphics[width=0.75\textwidth]{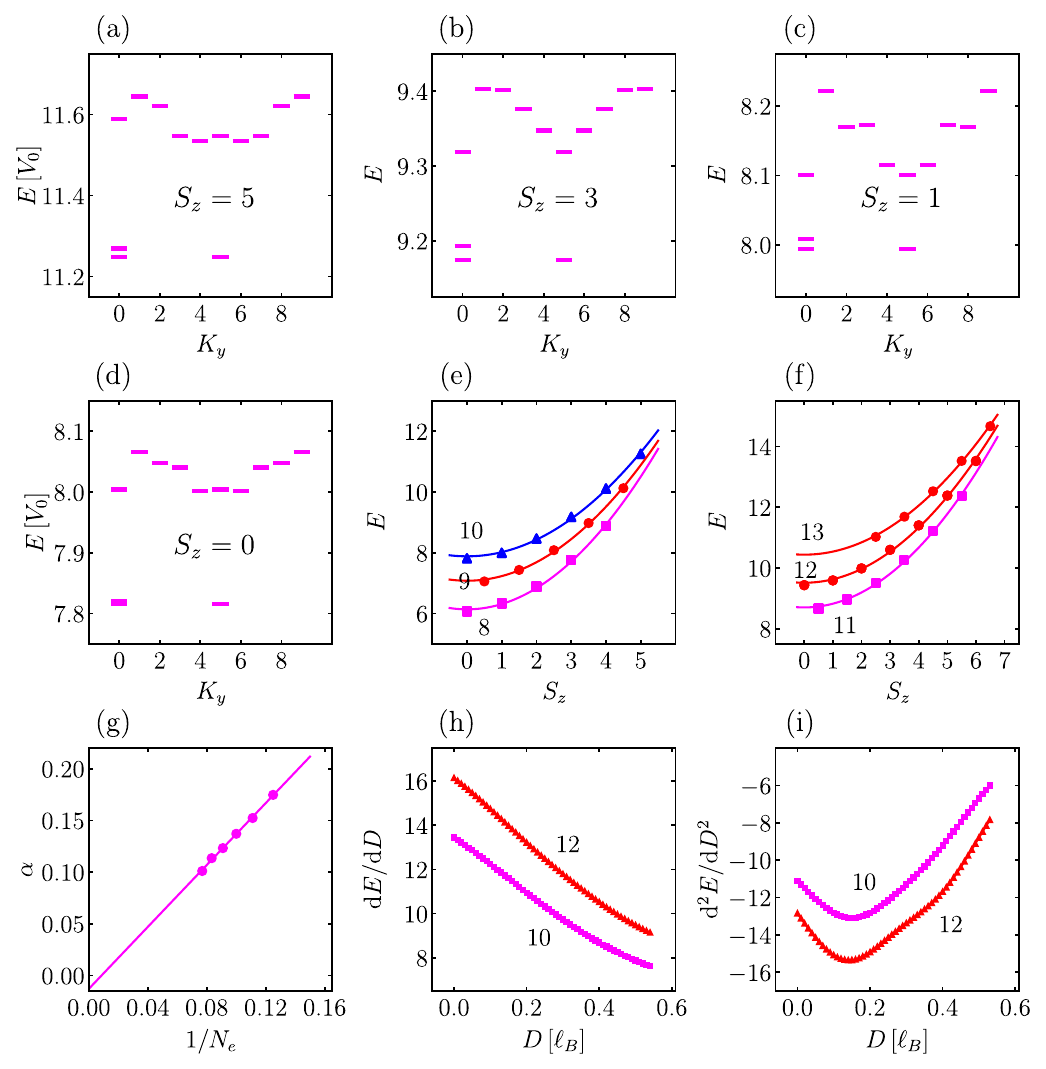}
\caption{(a-d) Energy spectra of the system with $N_{b}=10$ and $g=0.5$. The $S_{z}$ value is shown as inset in each panel. (e-f) Fitting of the modified eigenvalues $\widetilde{E}(S_{z})$ at $g=0.5$ using Eq.~\eqref{EnergyFormula}. The number of bosons for each curve is indicated. (g) Finite-size scaling of the coefficient $\alpha$ in Eq.~\eqref{EnergyFormula} at $g=0.5$. (h-i) First-order and second-order derivatives of the ground state energy with respect to $g$. The average of the eigenvalues of the quasi-degenerate ground states is used.}
\label{FigureS1}
\end{figure}

Numerical results for $g=0.5$ are presented in Fig.~\ref{FigureS1} and they are very similar to those in Fig.~\ref{Figure1}. We denote the number of bosons as $N_{b}$. Three quasi-degenerate ground states are found in each $S_{z}$ sector. The eigenvalue versus $S_{z}$ relation can also be fitted using Eq.~\eqref{EnergyFormula}. We have only five data points for $N_{b}=13$ due to large computational cost. For the short-range Hamiltonian, there is no capacitance energy when $S_{z}$ changes. The coefficient $\alpha$ scales to zero as $1/N_{b}$. These results demonstrate that the bosonic NAIC state is realized. Next we turn to the phase transition induced by tuning the interlayer coupling strength $g$. If both layers have filling factor $1/2$ and $g$ is sufficient small, the system simply consists of two decoupled bosonic Laughlin states. The system undergoes a transition at finite $g$ such that the NAIC state emerges. As in the fermionic cases, the ground state energy is defined as the average of the three quasi-degenerate ground states. The first and second-order derivatives of this average are shown in Fig.~\ref{FigureS1} (h-i). We may tentatively claim that the transition is continuous. It is a close analog of the transition between the fermionic NAIC state and Halperin 331 state. When real space wave functions are used to describe FQH systems, a bosonic wave function can be obtained from a fermionic one if it is divided by the Jastrow factor. This relation holds between the bosonic and fermionic NAIC state as well as the decoupled Laughlin states and the Halperin 331 state. There is a simple picture for the gapped sector of the transition in bosonic systems. While the Laughlin state is usually described by the U(1)$_{2}$ Abelian Chern-Simons theory, it can be recast as the SU(2)$_{1}$ non-Abelian Chern-Simons theory using level-rank duality. For one-component bosonic systems, the Moore-Read state is described by the SU(2)$_{2}$ non-Abelian Chern-Simons theory~\cite{Fradkin1998}. In general, the SU(2)$_{k}$ non-Abelian Chern-Simons theory has the Lagrangian density 
\begin{eqnarray}
\mathcal{L}_{k} = \frac{k}{4\pi} {\rm Tr} \left[ a{\wedge}da + \frac{2}{3} a{\wedge}a{\wedge}a \right],
\end{eqnarray}
where the gauge field $a$ takes values in the su(2) Lie algebra~\cite{Nayak2008}. On the side of decoupled Laughlin states, two fields $a_{1}$ and $a_{2}$ should be introduced. If they are locked together ($a_{1} \equiv a_{2}$) using a Higgs type mechanism~\cite{WuYH2023}, their Lagrangian density simply adds up to become that of the Moore-Read state. It is still unclear how the gapless sector emerges in the transition.  

\end{document}